\documentclass{article}

\usepackage[english]{babel}

\usepackage[letterpaper,top=2cm,bottom=2cm,left=3cm,right=3cm,marginparwidth=1.75cm]{geometry}

\usepackage{amsmath,amsfonts,amssymb}
\usepackage{algorithm}
\usepackage{algpseudocode}
\usepackage{graphicx}
\usepackage[colorlinks=true, allcolors=blue]{hyperref}
\usepackage{subcaption}
\usepackage{authblk}
\usepackage{xcolor}
\usepackage[
  backend      = biber,
]{biblatex}

\graphicspath{{../}}

\title{A continuous data assimilation closure for modeling statistically steady turbulence in large-eddy simulation}

\author[1]{Sagy R. Ephrati\footnote{Corresponding author (sagy@chalmers.se)}}
\author[2]{Arnout Franken}
\author[2]{Erwin Luesink}
\author[2]{Paolo Cifani}
\author[2,3]{Bernard J. Geurts}
\affil[1]{Department of Mathematical Sciences, Chalmers University of Technology and University of Gothenburg, 412 96 Gothenburg, Sweden}
\affil[2]{Mathematics of Multiscale Modeling and Simulation, Faculty EEMCS, University of Twente, 7500 AE Enschede,
The Netherlands}
\affil[3]{Multiscale Energy Physics, CCER, Faculty Applied Physics, Eindhoven University of Technology,
5600 MB Eindhoven, The Netherlands}

\date{}

\begin{document}
\maketitle
\begin{abstract}
%A closure model for large-eddy simulation (LES) based on the 3D-Var data assimilation method is presented. 
A closure model is presented for large-eddy simulation (LES) based on the three-dimensional variational data assimilation algorithm.
The approach aims at reconstructing high-fidelity kinetic energy spectra in coarse numerical simulations by including feedback control to represent unresolved dynamics interactions in the flow as stochastic processes. The forcing uses statistics obtained from offline high-fidelity data and requires only few parameters compared to the number of degrees of freedom of LES. This modeling strategy is applied to geostrophic turbulence on the sphere and enables simulating indefinitely at reduced costs. It proves to accurately recover the energy spectra and the zonal velocity profiles in the coarse model, for three generic situations.
\end{abstract}

\section{Introduction}
The high computational costs involved in fully resolved turbulent flow simulations form a major challenge for computational methods. The complexity of direct numerical simulations has prompted the development of simulation strategies that require significantly fewer computational resources. Among these is large-eddy simulation (LES) in which a filtered description of the dynamics determines that only the largest, most energetic scales of motion are resolved and an LES (or subfilter-scale) model is introduced to account for unresolved dynamics and discretization error \cite{geurts2022direct, geurts2003regularization, sagaut2005large}.
Recently, data-driven LES has become an active research field that focuses on using any available data of the flow to specify models for accurate coarse-grid flow simulations. Machine learning is commonly used, which has successfully reduced the computational cost while producing relevant results in various settings. Examples include computing a variable eddy viscosity \cite{beck2019deep} or subfilter-scale forces \cite{xie2020modeling}, approximating energy spectra \cite{kurz2023deep}, and specifying models minimizing the number of tunable parameters \cite{edeling2020reducing}. Despite these advances, a computational overall-best LES model has not yet been found. Instead, we propose combining 3D-variational data assimilation and LES to correctly nudge the evolution of coefficients in a spherical harmonics expansion such that the coarsened flow prediction closely matches reference statistics obtained from high-fidelity data.

An abstract ideal LES model was put forward by Langford and Moser \cite{langford1999optimal}, which minimizes instantaneous error in the resolved dynamics and yields exact agreement for spatial statistics. The derived model term is the average subfilter-scale contribution, conditioned to the current configuration of the resolved variables. Finding this conditional average is challenging in practice since each resolvable configuration corresponds to a distribution of fields with unresolvable small-scale dynamics. Nonetheless, this ideal model term may be approximated empirically when sufficient data of the resolved system is available. Estimation of this distribution and, in particular, its mean is what data assimilation is concerned with \cite{grewal2014kalman}. Data assimilation combines observations (data) with predictions to reduce uncertainties optimally \cite{majda2012filtering}. A Bayesian approach is commonly adopted to account for uncertainty and subsequently find a distribution of solutions in a probabilistic setting \cite{reich2015probabilistic, rosic2013parameter}. The mean of this distribution is the corresponding ideal LES model minimizing errors in the resolved dynamics. We consider the applicability of data assimilation algorithms in the context of LES closure and determine new sub-filter models using data assimilation theory. 

The goal of this paper is to employ ideas from data assimilation in the context of large-eddy simulation for geophysical applications. Specifically, we propose a method to correct model error in a statistical sense for fluid dynamical systems in a stationary state. The chaotic behavior of turbulent flows justifies reproducing flow statistics rather than the actual flow itself. To this end, we present a data-driven stochastic forcing technique and apply this to coarse numerical simulations of geostrophic turbulence on a sphere. The forcing stems from the 3D-Var data assimilation algorithm \cite{courtier1998ecmwf, blomker2013accuracy} applied to the spectral representation of the solution, of which the measured statistics are assimilated into the solution in a coarse numerical simulation. The result is a method similar to Fourier domain Kalman filtering \cite{majda2012filtering} and continuous data assimilation \cite{azouani2014continuous, gesho2016computational}. The forcing aims to reconstruct the time-averaged energy spectrum obtained from high-fidelity data, which is a necessary criterion for accurate coarse numerical simulations. The method thereby enables performing computationally cheap numerical simulations that retain key flow statistics while being able to simulate for an indefinite time. This can be used, e.g., for inexpensively generating accurate ensemble forecasts. The current study aims to aid the development of efficient simulation strategies by replacing high resolution with stochastic forcing terms when coherent spatial patterns are contained in the resolved flow \cite{holm2015variational, memin2014fluid}. Such approaches have found meaningful applications in recent studies of idealized ocean models, focusing on subgrid-scale modeling \cite{resseguier2017geophysicalI, frederiksen2006dynamical, frederiksen2017stochastic}, uncertainty quantification \cite{cotter2019numerically, resseguier2017geophysicalII, ephrati2023data}, and data assimilation \cite{cotter2020particle}.

Considerable effort has been made in recent years concerning correcting model error using data assimilation techniques and data-driven approaches. In \cite{levine2022framework}, the authors study the model errors of mechanistic methods to purely data-driven methods and find that hybrid approaches outperform purely data-driven methods. Examples of hybrid methods are closure models for LES and multiscale problems, for which machine learning has been applied frequently  \cite{sanderse2024scientific}.
Alternatively, model errors can be controlled by data assimilation. For example, in \cite{bach2023multi} the model error is minimized by adaptively weighing predictions based on their accuracy. Alternatively, model error may be reduced by assimilating statistics into a dynamical system, as recently proposed by \cite{bach2024filtering}. Data assimilation has also been applied to LES to optimize model parameters. The work \cite{mons2021ensemble} adopts an ensemble-variational data assimilation approach to optimize a Smagorinsky coefficient and a forcing term to match measured flow statistics. The study showed that the mean flow and Reynolds stresses could be accurately approximated, outperforming traditional LES methods. More recently, \cite{wang2023ensemble} employed a similar approach, optimizing the parameters in a mixed LES closure model to best approximate reference kinetic energy spectra. To our knowledge, there are no previous studies that develop LES closure models by continuously assimilating flow statistics into the dynamical system.

Data assimilation methods are used in the present study to determine the functional form of the data-driven LES closure model.  We consider flows that develop a statistically stationary state and derive a data-driven model that can be used online, during coarsened simulations. The proposed method can be specified entirely offline, exploiting already available data. After this preparatory phase, a stand-alone model is obtained with which the coarsened flow can be simulated indefinitely at reduced computational costs. This `offline-online' approach differs from so-called `continuous data assimilation' methods that employ measurements that become available sequentially in time during the simulations \cite{azouani2014continuous}.

The offline-online data assimilation approach considered here shows similarities with studies combining continuous data assimilation with reduced-order modeling. In these approaches, accurate coarse-grained results are achieved by combining numerical predictions with real-time data. A forcing term is then included in the prediction as it is integrated in time \cite{altaf2017downscaling, daley1983linear, charney1969use}. A recent example is available in \cite{zerfas2019continuous} where continuous data assimilation was used to improve the accuracy of reduced-order models of flow past a cylinder. In such an approach, data for assimilation is received and treated `on-the-fly', i.e., during the course of a simulation. In particular, \cite{zerfas2019continuous} investigates the effects of adding or removing dissipation from the reduced-order model. This approach can be used to control the kinetic energy of the flow and nudge this in the desired direction. This methodology was found to improve long-time accuracy. Continuous data assimilation is a preferred method when time-accurate coarsened predictions are sought. In this paper, the chosen approach of collecting and processing all data {\it{a priori}}, instead of on-the-fly, is suitable for data-driven LES of steady turbulent flow. By working with the entire dataset that is available from high-fidelity observations or simulations, one may incorporate also long-term flow characteristics and optimize the prediction of particular features such as the spectrum of the turbulence. This sets the currently presented method apart from continuous data assimilation. 

The technique presented in this paper has been applied to the two-dimensional Euler equations on the sphere \cite{ephrati2023qeuler} and two-dimensional Rayleigh-B\'enard convection \cite{ephrati2023rb, ephrati2022stochastic}, where reference spectra could be accurately reproduced in coarse simulations. The former led to accurate and stable long-term dynamics. The latter resulted in accurate average heat flux in the domain and generalized well to a range of Rayleigh numbers centered around the reference value for which high-fidelity data was available. In this paper, we demonstrate that this simple model recovers qualitative features of reference simulations of geostrophic turbulence. We achieve this using a tailored model to reconstruct reference energy spectra in coarse numerical simulations. These spectra serve as a key statistic for the flow dynamics and their reconstruction establishes the feasibility of the proposed method. We note that other global quantities of interest may be reconstructed similarly using tailored basis functions \cite{edeling2020reducing}. This motivates further development of data-driven LES strategies using data assimilation methods that approximate ideal models in the sense of Langford and Moser \cite{langford1999optimal} and serves as a step toward nudging methods for isotropic two-dimensional turbulence and fully developed three-dimensional turbulence.

The paper is structured as follows. The continuous data assimilation closure model is introduced in Section \ref{sec:model_description} for dynamical systems in general. The application to coarse-grained geostrophic turbulence is highlighted in Section \ref{sec:application}, detailing the underlying equations in Section \ref{subsec:governing_eqs} and assessing model performance in Sections \ref{subsec:known_ic} to \ref{subsec:random_ic}. The conclusions are presented in Section \ref{sec:conclusions}.

\section{LES closure from the 3D-Var data assimilation algorithm}
\label{sec:model_description}
The continuous data assimilation closure model that we propose and test here is based on the premise that the average energy spectrum of the coarse numerical solution should equal that of the reference solution, up to the smallest resolvable scale on the coarse computational grid. That is, the measured energy spectrum is truncated at some coarse resolution $N$ and serves here as the key statistic to be reproduced in a coarse numerical simulation. This imposes a constraint on the coefficients of the spectral representation of the coarse numerical solution, leading to the model derivation via three steps outlined below.

The first step is a modal expansion of the dynamics. A (discretized) fluid dynamical system with prognostic variable $q$ described by \begin{equation}
\frac{\text{d}q}{\text{d}t} = L(q),
\label{eq:spectral_expansion_general}
\end{equation} 
for some operator $L(q)$ can be projected onto a suitable set of basis functions. A basis of spherical harmonics is adopted in the current study, which is a natural choice for flows on the sphere. Naturally, different domains necessitate the use of different basis functions. For example, a Fourier basis can be adopted on a periodic domain \cite{ephrati2023rb}, whereas a basis obtained from proper orthogonal decomposition (POD) is suited for general domains \cite{ephrati2022stochastic}. The current basis is denoted by $\{Y_{lm}\}$, where $l=0,\ldots, N-1$ denotes the degree of the spherical harmonic function. Here $N$ is the adopted resolution. A total of $2l+1$ basis functions exist for each degree $l$, denoted by the order $m =-l,\ldots, l$.  Projecting the solution onto the basis functions allows for retrieving the time-dependent coefficients $\{c_{lm}\}$, that express the solution in the selected basis. The model will act on the level of these coefficients. Their evolution is denoted by
\begin{equation}
\frac{\text{d}c_{lm}}{\text{d}t} = \langle L(q), Y_{lm}\rangle =: L_\mathrm{c}(\mathbf{c}, l, m)
\end{equation}  
where $\langle\,\cdot\,,\,\cdot\,\rangle$ is the spatial inner product, $c_{lm}$ is the expansion coefficient corresponding to $Y_{lm}$ and $\mathbf{c}$ is a vector containing all these coefficients. Similarly, the evolution of the magnitude of the coefficients is given by 
\begin{equation}
\frac{\text{d}|c_{lm}|}{\text{d}t} =: L_r(\mathbf{c}, l, m),
\label{eq:L_r}
\end{equation} 
for an associated operator $L_r$. Note that our approach does not require the actual form of $L_c$ or $L_r$, which in practice will also depend on the adopted discretization and resolution. Instead, we only require a transformation from the numerical solution in physical space to the expansion coefficients and vice versa. Respectively, these are defined as $c_{lm} = \langle q, Y_{lm}\rangle$ and $q = \sum_{l, m} c_{lm}Y_{lm}$. 

The second step is to formulate the model as stochastic forcing, to represent the unresolved dynamics and inherent incertainty. % \textcolor{red}{Sagy, je hebt nog niet gezegd waarom je naar zo'n type model zoekt - wat is het 'voordeel' van een stochastische forcering? Of kun je dit motiveren vanuit Langford 1999?}. 
It is observed that the average energy level corresponding to $c_{lm}$ is given by $\mathbb{E}(|c_{lm}|^2)$, which satisfies 
\begin{equation}
    \mathbb{E}\left(|c_{lm}|^2\right) = \mathrm{var}(|c_{lm}|) + \mathbb{E}\left(|c_{lm}|\right)^2
\end{equation} 
in a statistically stationary state. Thus an accurate energy spectrum can be obtained when achieving accurate mean values and variances of the magnitudes of the coefficients. We accomplish this by including a feedback control term in the evolution of the magnitudes as 
\begin{equation}
\text{d}|c_{lm}| = L_r(\mathbf{c}, l, m)\text{d}t + \frac{1}{\tau_{lm}}\left(\mu_{lm} - |c_{lm}|\right)\text{d}t + \sigma_{lm}\text{d}W_{lm},
\label{eq:nudging_continuous}
\end{equation}
where $\text{d}W_{lm}$ denotes Gaussian noise. The additional Ornstein-Uhlenbeck (OU) process arises in the continuous-time limit of the 3D-Var data assimilation algorithm \cite{blomker2013accuracy}. The noise term commonly appears in data assimilation to emulate noisy observations, although the original formulation of the continuous data assimilation method \cite{azouani2014continuous} is deterministic. In the current approach, the noise term is included to realize an accurate reproduction of the reference variance. Numerical simulations without a noise term were also studied in previous studies \cite{ephrati2023qeuler, ephrati2023rb}. The OU process in \eqref{eq:nudging_continuous} is here defined for each coefficient separately with mean $\mu_{lm}$, noise scaling $\sigma_{lm}$, and forcing strength determined by the time scale $\tau_{lm}$, which will be defined in the third step. These are discretized as a prediction-correction scheme, incorporating the feedback term via nudging (Newtonian relaxation). This is summarized as
\begin{eqnarray}
    \tilde{c}^{n+1}_{lm} = \int_{t^n}^{t^{n+1}} L_c(\mathbf{c}^n, l, m)\text{d}t, \label{eq:prediction}\\
    |c^{n+1}_{lm}| = |\tilde{c}^{n+1}_{lm}| + \frac{\Delta t}{\tau_{lm}}\left(\mu_{lm} - |\tilde{c}^{n+1}_{lm}|\right) + \sigma_{lm}\Delta W_{lm}^{n+1}, \label{eq:correction}
\end{eqnarray}
where the superscripts indicate the time instances and $\Delta W_{lm}^{n+1}$ is drawn from a standard normal distribution. The correction is independent of the time-integration method in \eqref{eq:prediction}, which is not required to be in spectral space. The correction \eqref{eq:correction} acts only on the magnitude of the basis coefficients.% while the phases remain unaltered as they do not affect the energy spectrum.

The third and final step of the model specification concerns the definition of the forcing parameters. If a reference mean value and variance for $|c_{lm}|$ are known, then any stochastic process that models $\text{d}c_{lm}$ and reproduces this mean and variance will recover the desired energy level for $c_{lm}$. Below, we elaborate on a stochastic process that includes $L_r$, and thus the underlying physics, and simultaneously approximates the mentioned statistics. This implies that the energy spectrum may be reconstructed while incorporating the original dynamics of the governing equations. In the 3D-Var algorithm, the evolution operators $L_c$ and $L_r$ are treated as the identity operators \cite{lorenc2005does}. This assumption is appropriate in statistically stationary states and for sufficiently small time step sizes. Under these assumptions, the evolution of $|c_{lm}|$ in \eqref{eq:prediction}-\eqref{eq:correction} can be treated as the first-order autoregressive (AR(1)) process with mean $\mu_{lm}$, drift coefficient $(1-\Delta t/ \tau_{lm})$ and noise variance $\sigma^2_{lm}$. We assume that high-fidelity snapshots are available a priori from which the reference mean $\mathbb{E}(|{c_{lm, \mathrm{ref}}}|)$ and variance $\mathrm{var}(|c_{lm, \mathrm{ref}}|)$ are extracted. In the present study these snapshots are collected from a high-resolution simulation, as a synthetic substitute for observational data which might be used in, e.g., numerical weather prediction. To actually attain the measured reference values in the AR(1) process, we require that $\mu_{lm}=\mathbb{E}(|{c_{lm, \mathrm{ref}}}|)$ and \begin{eqnarray}
    \sigma_{lm} = \sqrt{\mathrm{var}(|c_{lm, \mathrm{ref}}|)}\sqrt{1 - \left(1 - \frac{\Delta t}{\tau_{lm}}\right)^2}.
\end{eqnarray}
This leaves $\tau_{lm}$ as the only free parameter, drastically reducing the number of degrees of freedom of the model. Here, we choose $\tau_{lm}$ heuristically as the measured correlation time of the high-fidelity time series of $|c_{lm, \mathrm{ref}}|$. The AR(1) process becomes Gaussian noise in the limiting case of $\tau_{lm}\leq\Delta t$ and becomes deterministic in the limit of large $\tau_{lm}$. Assuming, as in 3D-Var, that $L_c$ and $L_r$ can be regarded as identity operators, the acquired model parameters are obtained independent of the adopted discretization and coarse resolution. Within this approximation the model parameters only depend on the high-fidelity data. While the reference high-resolution kinetic energy spectrum defines the forcing, the forcing by itself does not prescribe the energy levels in the coarse numerical simulations but only contributes to their dynamics. The combination of both the coarse discretization and the forcing terms extracted from the data determines how the spectral coefficients adapt in time.

The prediction-correction procedure (\ref{eq:prediction}-\ref{eq:correction}) is summarized in Algorithm \ref{alg:cap}. Here, we assume that the values for $\mu_{lm}, \sigma_{lm}$, and $\tau_{lm}$ are known and that a vorticity field $q^n$ at time $t^n$ is available. The summation over all available modes in Algorithm \ref{alg:cap} can be replaced by a fraction of the modes to reduce the range of basis function at which the forcing is applied. This is demonstrated in Section \ref{subsec:partial_forcing}.
\begin{algorithm}
    \caption{Prediction-correction scheme (\ref{eq:prediction}-\ref{eq:correction}) for one time step}\label{alg:cap}
    \begin{algorithmic}
        \Procedure{Predictor-Corrector}{$q^n, L, \Delta t, \mu_{lm}, \sigma_{lm}, \tau_{lm}$}
        \State $\tilde{q}^{n+1} \gets \int_{t^n}^{t^{n+1}} L(q^n)\,\text{d}t$ \Comment{Time integration / prediction}
        \For{$l=0,\ldots, N-1$}
            \For{$m=-l,\ldots, l$}
                \State $\tilde{c}_{lm}^{n+1} \gets \langle \tilde{q}^{n+1}, Y_{lm} \rangle$ \Comment{Projection onto basis vector}
                \State $r \gets \mathrm{abs}(\tilde{c}_{lm}^{n+1})$
                \State $\phi \gets \mathrm{angle}(\tilde{c}_{lm}^{n+1})$
                \State $\Delta W \gets$ sample from $\mathcal{N}(0,1)$
                \State $r \gets r + \frac{\Delta t}{\tau_{lm}}\left(\mu_{lm} - r\right) + \sigma_{lm}\Delta W$ \Comment{Correction of the magnitude}
                \State $c_{lm}^{n+1}\gets r\cdot\exp{i\phi}$ \Comment{Reconstruction of the basis coefficient}
            \EndFor
        \EndFor
        \State $q^{n+1} = \sum_{l=0}^{N-1}\sum_{m=-l}^l c_{lm}^{n+1}Y_{lm}$ \Comment{Reconstruction of the vorticity field}
        \State \Return $q^{n+1}$
        \EndProcedure
    \end{algorithmic}
\end{algorithm}

The prediction-correction procedure (\ref{eq:prediction}-\ref{eq:correction}) can be placed in the context of data assimilation by defining a prediction and an observation. The prediction is obtained by evolving the prognostic variable according to the coarse-grid discretization. Subsequently, the `observations' are flow fields with the desired energy spectrum, which depend only on the means $\mu_{lm}$ and variances $\sigma^2_{lm}$. In this light, the nudging approach \eqref{eq:correction} acts as a correction and can be understood as a steady-state Kalman-Bucy filter \cite{grewal2014kalman} with fixed gain $\Delta t / \tau_{lm}$.  In total, this yields a method that relies both on the coarse discretization and on the data. The unresolved interactions between the expansion coefficients are modeled as linear stochastic processes, which also underlies Fourier domain Kalman filtering \cite{harlim2008filtering, majda2012filtering}. In the offline-online approach, all reference data is collected independently first, and the model parameters are determined before any coarse numerical simulations are performed. This corresponds to the approach typically embraced in data-driven LES \cite{beck2019deep}.

\section{Application to geostrophic turbulence}
\label{sec:application}
In this section, we demonstrate the continuous data assimilation closure for LES by applying it to the rotating Navier-Stokes equations (RNSE) and the quasi-geostrophic equations (QGE). The governing equations and adopted numerical methods are introduced in Section \ref{subsec:governing_eqs}. The model is subsequently assessed in a series of numerical experiments. The first results deal with predictions initialized from the reference steady state while forcing at all modes, in Section \ref{subsec:known_ic}, and only a part of the modes, in Section \ref{subsec:partial_forcing}. Section \ref{subsec:random_ic} concerns the model performance using random initial conditions.

\subsection{Governing equations and numerical methods}
\label{subsec:governing_eqs}
The RNSE and QGE are part of a larger family of geophysical fluid dynamical models \cite{holm2021stochastic} and are cornerstones in the study of rotating fluids on a planetary scale. In terms of potential vorticity $q$ and stream function $\psi$, the QGE on the sphere read \cite{franken2023zeitlin} 
\begin{eqnarray}
    \dot{q} = \left\{ \psi, q \right\} + \nu\left(\Delta \omega + 2 \omega \right) - \alpha \omega + f, \\
    \left(\Delta - \gamma\mu^2\right) \psi = \omega, \\
    \omega = q - 2\mu.
\end{eqnarray}

Here, the Poisson bracket $\left\{\psi, q\right\}$ governs the advection of $q$, $\nu$ is the viscosity, $\alpha$ the Rayleigh friction and $f$ an external forcing. Nonlinear Coriolis effects are included via $\mu = \sin\phi$ with $\phi$ the latitude on the sphere, while $\gamma$ denotes the Lamb parameter~\cite{vallis2019essentials}. The Lamb parameter is determined by the radius of the sphere $R$ and the Rossby deformation length $R_d$, \begin{equation}
    \gamma = 4\frac{R^2}{R_d^2}, \quad \text{ where } R_d = \frac{\sqrt{gH}}{\Omega}.
\end{equation}
Here, $g$ is the gravitational acceleration, $H$ is the average fluid layer thickness, and $\Omega$ is the rotation frequency of the sphere.
We denote the longitude by $\theta$. Without loss of generality, we set the radius of the sphere to unity, for which the Poisson bracket can be written in coordinates as \begin{equation}
    \left\{\psi, q\right\}(\phi,\theta) = \frac{1}{\cos \phi} \left(\frac{\partial \psi}{\partial \phi}\frac{\partial q}{\partial \theta} - \frac{\partial \psi}{\partial \theta}\frac{\partial q}{\partial \phi}\right).
\end{equation}
We refer to \cite{luesink2024geometric} for a comprehensive derivation of the QGE on the sphere.

For the quasi-geostrophic equations, the evolution of the spectral coefficients \eqref{eq:spectral_expansion_general} can be further expanded as
\begin{equation}
    \begin{split}
        L_\mathrm{c}(\mathbf{c}, l, m) &= \langle \left\{\psi, q\right\}, Y_{lm}\rangle +
        \langle \nu(\Delta \omega + 2\omega), Y_{lm}\rangle
        - \langle \alpha \omega , Y_{lm}\rangle
        + \langle f, Y_{lm}\rangle \\
        &= \langle \left\{\psi, q\right\}, Y_{lm}\rangle + \left(-l(l+1)\nu + 2\nu-\alpha\right)c_{lm} + \langle f, Y_{lm} \rangle.
    \end{split}
    \label{eq:spectral_expansion}
\end{equation}
The last term on the right-hand side of \eqref{eq:spectral_expansion} depends on the external forcing. In the high-resolution simulations of the QGE performed by \cite{franken2023zeitlin}, the forcing is localized in a narrow band around degree $l=100$ and will therefore lead only to a nonzero contribution of $\langle f, Y_{lm}\rangle$ for the spectral coefficients within this band. Further expanding the transport term $\langle \left\{\psi, q\right\}, Y_{lm}\rangle$ leads to a complicated expression in which the transport spectrum can be expressed as a bilinear form on the spectra of $\psi$ and $\omega$ \cite{platzman1960spectral}. The triadic interactions of the spherical harmonic modes in the transport term are rigorously analyzed in \cite{platzman1960spectral}. In particular, the most stringent criterion for interaction between different wavenumbers $l_1, l_2, l_3$ is that $l_3 = l_2 + l_1$. 
We note, however, that further expansion of the evolution of the coefficients ($L_c$) or their magnitudes ($L_r$ in Eq. \ref{eq:L_r}) is not necessary for applying the proposed closure model. Instead, only the transformations from the vorticity field to the spectral coefficients and vice versa are required.

In the absence of dissipation and external forcing and damping, the dynamics are fully governed by the convective term $\{\psi, q\}$. In these cases, the functions given by the integrated powers of potential vorticity \begin{equation}
    \mathcal{C}_k(q) = \int_{S^2}\! q^k \,\text{d}x
\end{equation}
are conserved quantities and are referred to as Casimir functions. 

The governing equations are discretized using the spatial Zeitlin discretization \cite{zeitlin2004self}, which is suitable for the Navier-Stokes equations \cite{cifani2023efficient} and QGE \cite{franken2023zeitlin} on the sphere. Time integration is performed as reported by \cite{cifani2022casimir}, using second-order Strang splitting \cite{strang1968construction} where viscous dissipation, external forcing and damping are treated via a Crank-Nicolson scheme \cite{crank1947practical} and the convective term is integrated with a Casimir-preserving time integrator \cite{modin2020casimir}. Combined, these methods provide a second-order accurate discretization of the dynamics while conserving integrated powers of vorticity in the absence of external forcing and viscosity. In particular, this implies the conservation of the total vorticity and the enstrophy (integrated squared vorticity). The high-fidelity numerical experiments used here are originally presented by Franken et. al \cite{franken2023zeitlin}. The reference data are obtained at high resolution $(N=1024)$ with external forcing applied at wavenumber $l=100$ to ensure that a nontrivial statistical steady state is reached.

We consider three test cases representing distinct flow regimes by varying the Lamb parameter $\gamma$. We adopt $\gamma=0$, recovering the RNSE, $\gamma=10^3$, and $\gamma=10^4$. These values are based on physical parameters relevant for Earth applications \cite{schubert2009shallow}. The adopted Rayleigh friction constant $\alpha$ is chosen as $2\times 10^{-2}$, which was found to avoid an accumulation of energy in the largest flow scales whilst balancing the energy injected by the external forcing \cite{cifani2022casimir}. The dissipation $\nu$ is set at $10^{-6}$ in dimensionless units. The viscosity should not be regarded as molecular viscosity but instead represents the sub-grid enstrophy dissipation \cite{maltrud1993energy}.

Including enstrophy dissipation ensures that the flow is fully resolved at the chosen resolution. A double cascade in the energy spectrum is observed once the statistically steady state is reached, in agreement with known theoretical and experimental results on two-dimensional turbulence \cite{kraichnan1967inertial, lindborg2022two, cifani2022casimir}. This is visible in the non-zonal modes, which follow scaling laws of $-5/3$ and $-3$. Zonal jets are formed in the solutions due to the rotation of the sphere, leading to zonal modes dominating the larger scales of motion \cite{franken2023zeitlin}.

The characteristic length scale is defined as the radius of the sphere and leads to a Rossby number $\mathrm{Ro}=10^{-3}$ for the adopted rotation speed. This value is similar to the Rossby number for oceanic flows \cite{luesink2024geometric}. The characteristic time is defined as the rotational period of the sphere, here chosen as $\Omega^{-1}=6\times10^{-3}$ time units. The characteristic velocity is determined by the maximal zonal velocity and is approximately $0.5$, expressed as characteristic length per characteristic time. For these parameters, the Reynolds number $\mathrm{Re} \approx 5.3\times 10^5$ is found. It is important to bear in mind that the Reynolds number is not based on conventional molecular dissipation. The reference simulations and coarse simulations are carried out with a step size $\Delta t = 10^{-4}$ and $\Delta t = 5\times 10^{-3}$, respectively. These values correspond to approximately 60 and 30 time steps per characteristic time. All reference data is collected from 200 consecutive snapshots in the statistically steady state, each separated by $0.1$ time units.

The overall complexity of the employed numerical method is $\mathcal{O}(N^3)$ per time step for an adopted resolution $N$ \cite{cifani2023efficient}. A significant reduction in computational costs and memory requirements can therefore be achieved if accurate predictions are possible on coarse computational grids. The implementation of the closure term introduced in the previous section induces some overhead computational costs, primarily due to the conversion between the vorticity field and the spectral coefficients. Coarse resolutions of $N=24, 32, 48$ are considered in the next section. At these resolutions, the overhead costs were timed and typically amount to less than $12\%, 16\%,$ and $25\%$, respectively, of the runtime of an integration step. The closure yields predictions of relevant accuracy at very modest costs compared to the high-resolution computations. The simulation speed-up basically follows the $N^3$ scaling. Since we may successfully simulate the dominant dynamics on grids that are approximately 21 to 42 times coarser than the grid employed for the reference solution, the speed-up per time step amounts to a factor around $((1-0.25)\cdot21)^3\approx 3.9\times 10^{3}$ in the worst case to $((1-0.12)\cdot42)^3 \approx 5.0\times 10^4$ times in the best case. These estimates are obtained after all offline preparations were incorporated. Additionally, the coarse computational grids allow for a larger time step size than the value adopted on the reference grid, although this further decrease in computational cost was not considered in the presented cost estimates.
Other additional calculations, such as the computation of the model parameters from the data, take place during the offline step and do not affect the online computational efforts.

\subsection{Model predictions from initial conditions in the steady state}
\label{subsec:known_ic}
In what follows, we assess the closure model at coarse resolutions of $N=24, 32, 48$, i.e., much smaller than resolution $N=1024$ that is required for high-fidelity simulations. The external forcing is focused on wavenumber $l=100$, which cannot be resolved explicitly at the selected coarse levels of resolution. Hence, direct comparison between different resolutions would be meaningless without explicit modeling. This is illustrated by comparing results obtained with and without the model while initializing the simulations from a filtered high-resolution snapshot. Furthermore, this emphasizes the ability of the model to statistically correct errors arising from unresolvable dynamics on coarse grids. The coarse simulation results are compared qualitatively via instantaneous potential vorticity snapshots and quantitatively through the average energy spectrum, the vorticity distribution, the zonal velocity, and contributions of each of the terms appearing in the dynamics \eqref{eq:spectral_expansion} per wavenumber $l$. The solutions are compared at a lead time of 40 time units, which corresponds to 8000 coarse-grid time steps or approximately 267 rotations of the sphere.
    
The contribution of each of the terms in the dynamics relates to the inter-scale energy transfer. Specifically, the nonlinear transport term $\left\{ \psi, q\right\}$ ensures interactions between different scales of motion and strongly influences the inter-scale energy transfer. The use of coarse computational grids affects this transfer because of the truncation of the dynamics to the resolvable scales \cite{thuburn2014cascades}. A correct transfer of energy between the largest resolved scales of the flow is ideally achieved by including a closure model \cite{geurts2003regularization, vreman1997large}. The rate of change of the energy per wavenumber $l$ generally depends on more than one spectral component and the corresponding phases. Correspondingly, it can be used as an independent measure to assess the model performance as this rate of change was not explicitly included in the design of the closure model.

\begin{figure}[h!]
    \centering
    \includegraphics[width=\textwidth]{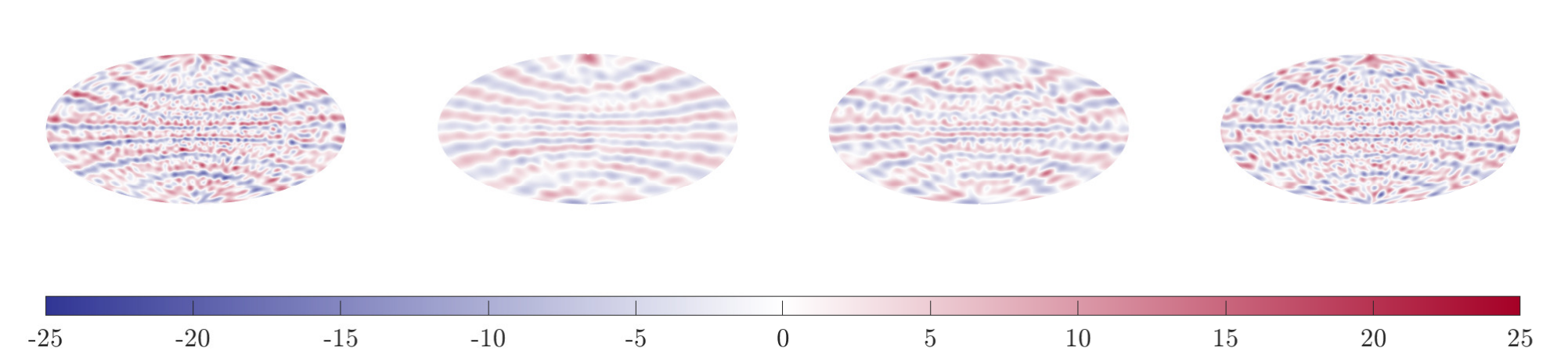}
    \caption{Instantaneous vorticity snapshots for the RNSE initialized from a filtered high-resolution snapshot in the statistically stationary state. Left: reference solution ($N=1024$), displaying only modes resolvable for $N=48$ for a qualitative comparison with the coarse model results. The solutions at coarse computational grids are obtained by applying the closure model at $N=24$ (middle left), $N=32$ (middle right), and $N=48$ (right). The vorticity fields are obtained 40 time units after initializing from the reference statistically stationary state.}
    \label{fig:lt_NS_vort_snaps}
    \centering
    \includegraphics[width=\textwidth]{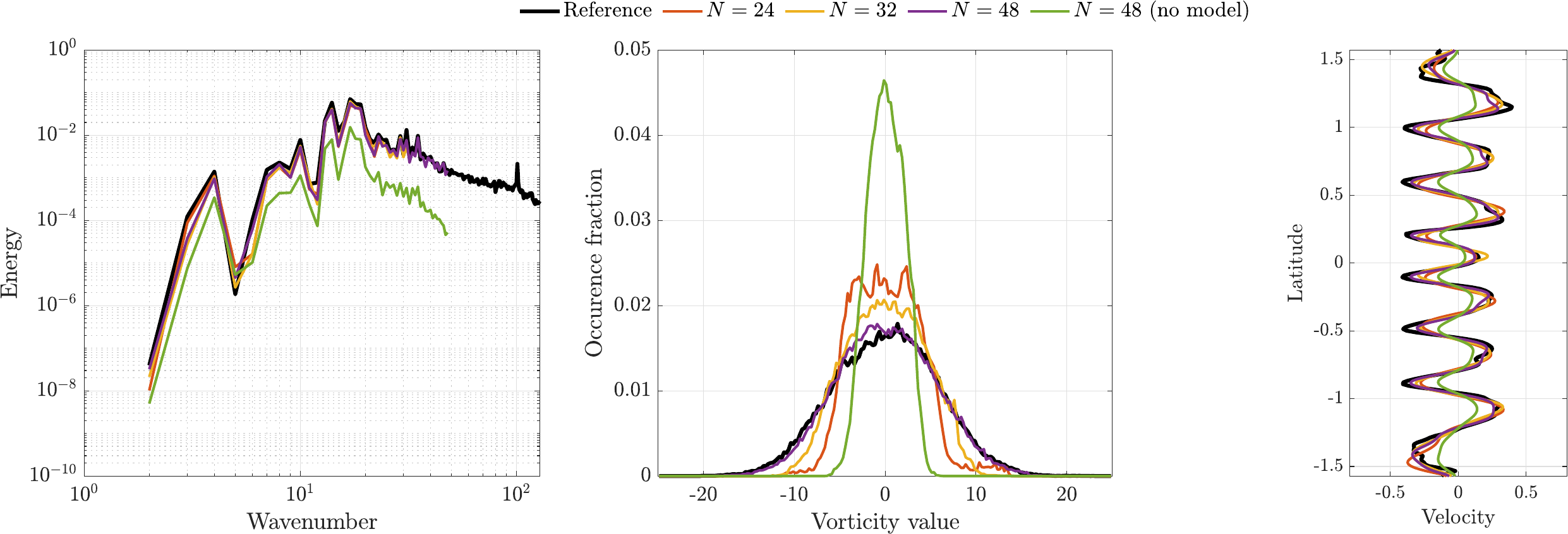}
    \caption{Instantaneous energy spectra (left), vorticity distribution (center), and zonal velocity (right), for the RNSE. The vorticity fields are obtained 40 time units after initializing from the reference statistically stationary state.}
    \label{fig:lt_NS_vort_info}
    \centering
    \includegraphics[width=\textwidth]{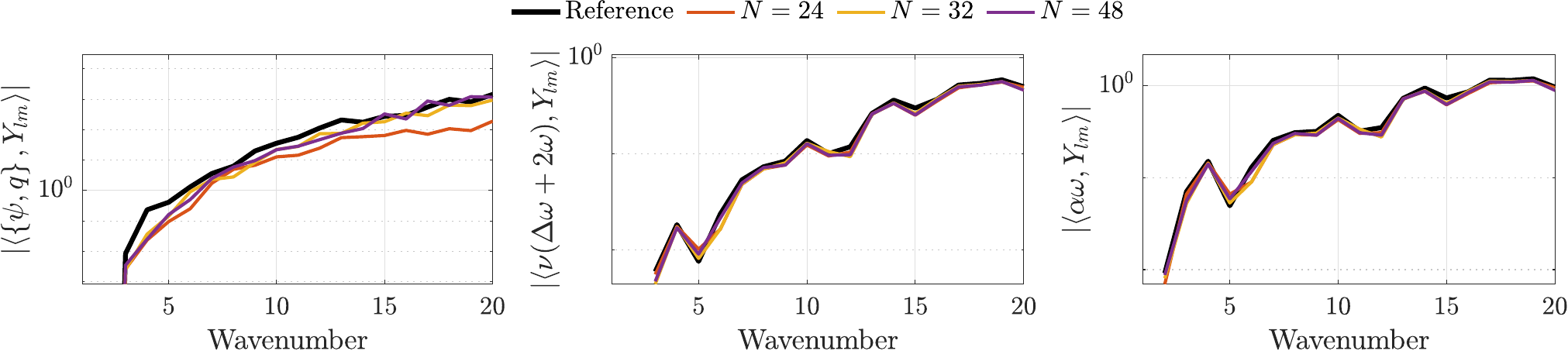}
    \caption{Instantaneous contributions per wavenumber for each of the terms in \eqref{eq:spectral_expansion} for the RNSE. Shown are the contributions per time unit of the convective term (left), the diffusion terms (center), and the friction term (right). The quantities are measured 40 time units after initializing from the reference statistically stationary state.}
    \label{fig:lt_NS_transfers}
\end{figure}

The results for the RNSE are shown in Figures \ref{fig:lt_NS_vort_snaps} to \ref{fig:lt_NS_transfers}. The instantaneous vorticity snapshots in Figure \ref{fig:lt_NS_vort_snaps} reveal that a good agreement of the zonal structures in the vorticity field is obtained with the model and demonstrate that increasing the resolution of the coarse numerical simulations yields instantaneous vorticity fields with increasingly smaller features. By construction of the forcing method, the small-scale features comply with the desired kinetic energy, as is observed in Figure \ref{fig:lt_NS_vort_info}. This establishes that the continuous data assimilation closure model presented in Section \ref{sec:model_description} indeed improves the energy spectra of the prognostic variable. This contrasts with the no-model results, where the energy decreases throughout the simulation due to viscosity and damping, as may also be seen in the instantaneous vorticity distributions and the zonal velocity. The loss of energy in the no-model simulation results in a vorticity distribution concentrated around zero and a decreased zonal velocity. The individual contributions to the dynamics of each term in \eqref{eq:spectral_expansion_general} are shown in Figure \ref{fig:lt_NS_transfers}. These results indicate that the difference between the model results and the reference is primarily caused by differences in the energy transfer due to convection.

\begin{figure}[h!]
    \centering
    \includegraphics[width=\textwidth]{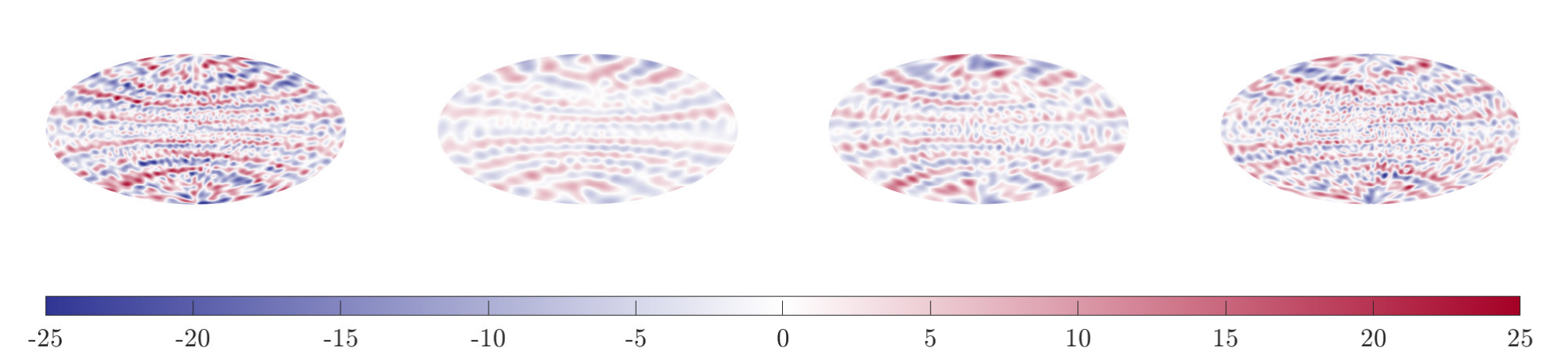}
    \caption{Instantaneous vorticity snapshots for the QGE with Lamb parameter $\gamma=10^3$ initialized from a filtered high-resolution snapshot in the statistically stationary state. Left: reference solution ($N=1024$), displaying only modes resolvable for $N=48$ for a qualitative comparison with the coarse model results. The solutions at coarse computational grids are obtained by applying the closure model at $N=24$ (middle left), $N=32$ (middle right), and $N=48$ (right). The vorticity fields are obtained 40 time units after initializing from the reference statistically stationary state.}
    \label{fig:lt_QG_vort_snaps}
    \centering
    \includegraphics[width=\textwidth]{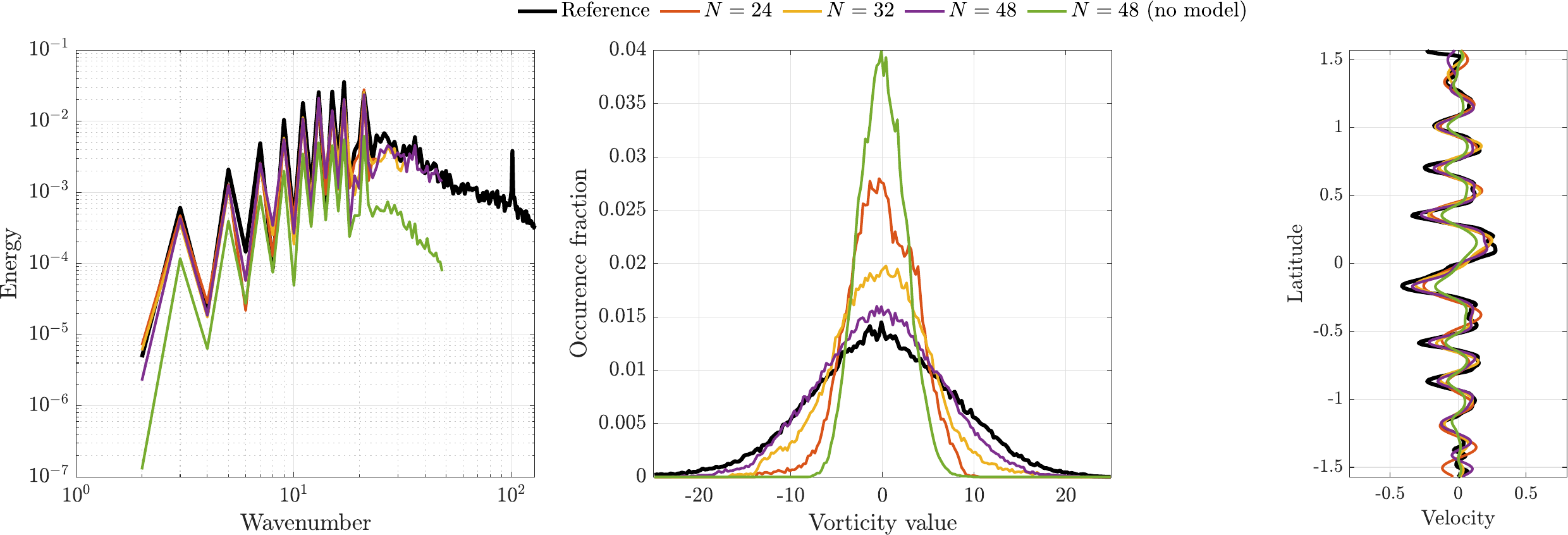}
    \caption{Instantaneous energy spectra (left), vorticity distribution (center), and zonal velocity (right), for the QGE with Lamb parameter $\gamma=10^3$. The vorticity fields are obtained 40 time units after initializing from the reference statistically stationary state.}
    \label{fig:lt_QG_vort_info}
    \centering
    \includegraphics[width=\textwidth]{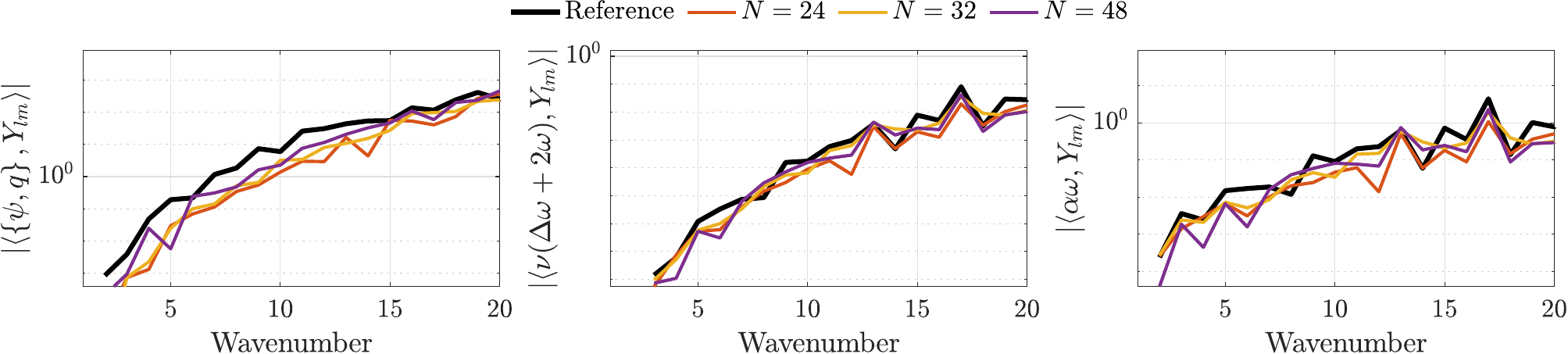}
    \caption{Instantaneous contributions per wavenumber for each of the terms in \eqref{eq:spectral_expansion} for the QG with Lamb parameter $\gamma=10^3$. Shown are the contributions per time unit of the convective term (left), the diffusion terms (center), and the friction term (right). The quantities are measured 40 time units after initializing from the reference statistically stationary state.}
    \label{fig:lt_QG_transfers}
\end{figure}

\begin{figure}[h!]
    \centering
    \includegraphics[width=\textwidth]{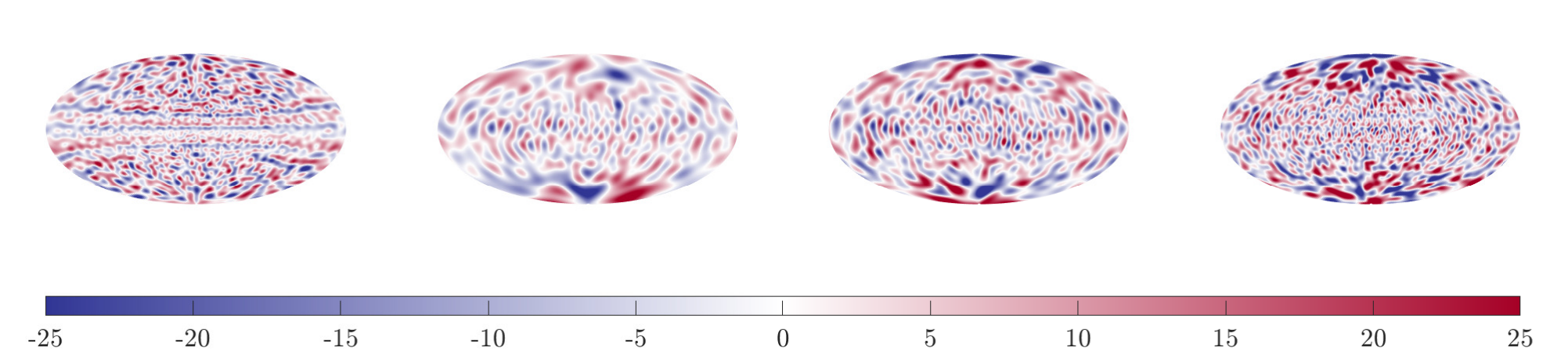}
    \caption{Instantaneous vorticity snapshots for the QGE with Lamb parameter $\gamma=10^4$ initialized from a filtered high-resolution snapshot in the statistically stationary state. Left: reference solution ($N=1024$), displaying only modes resolvable for $N=48$ for a qualitative comparison with the coarse model results. The solutions at coarse computational grids are obtained by applying the closure model at $N=24$ (middle left), $N=32$ (middle right), and $N=48$ (right). The vorticity fields are obtained 40 time units after initializing from the reference statistically stationary state.}
    \label{fig:lt_QG2_vort_snaps}
    \centering
    \includegraphics[width=\textwidth]{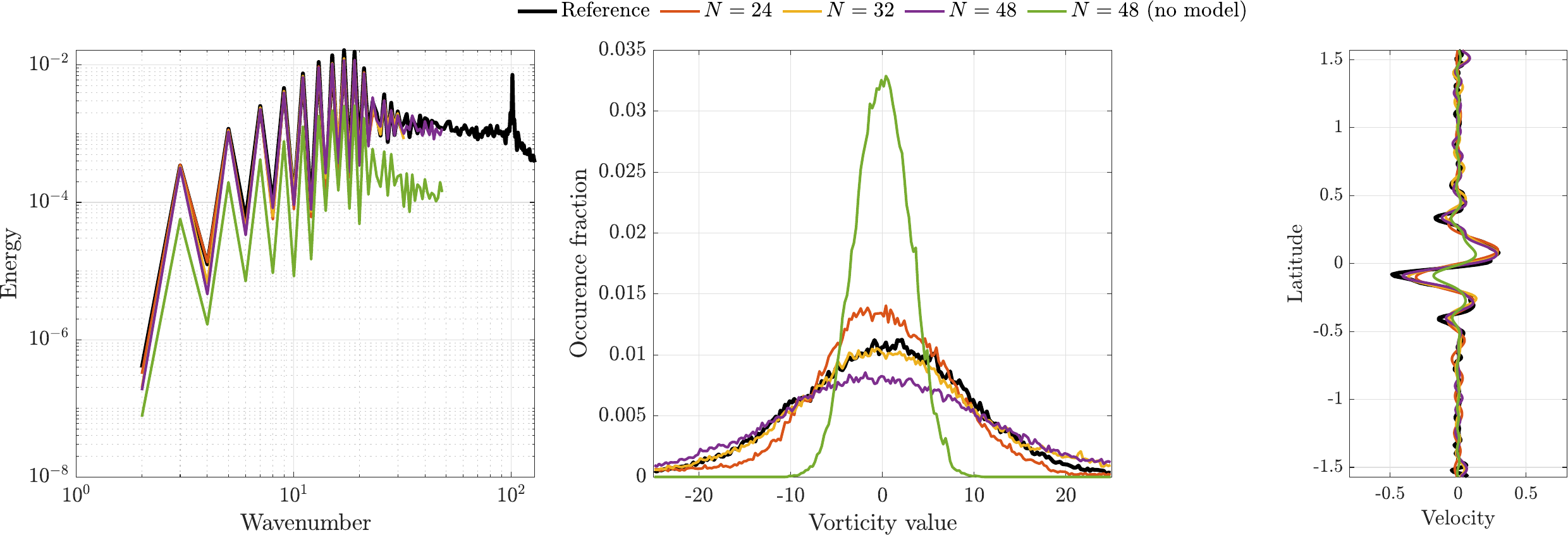}
    \caption{Instantaneous energy spectra (left), vorticity distribution (center), and zonal velocity (right), for the QGE with Lamb parameter $\gamma=10^4$. The vorticity fields are obtained 40 time units after initializing from the reference statistically stationary state. }
    \label{fig:lt_QG2_vort_info}
    \centering
    \includegraphics[width=\textwidth]{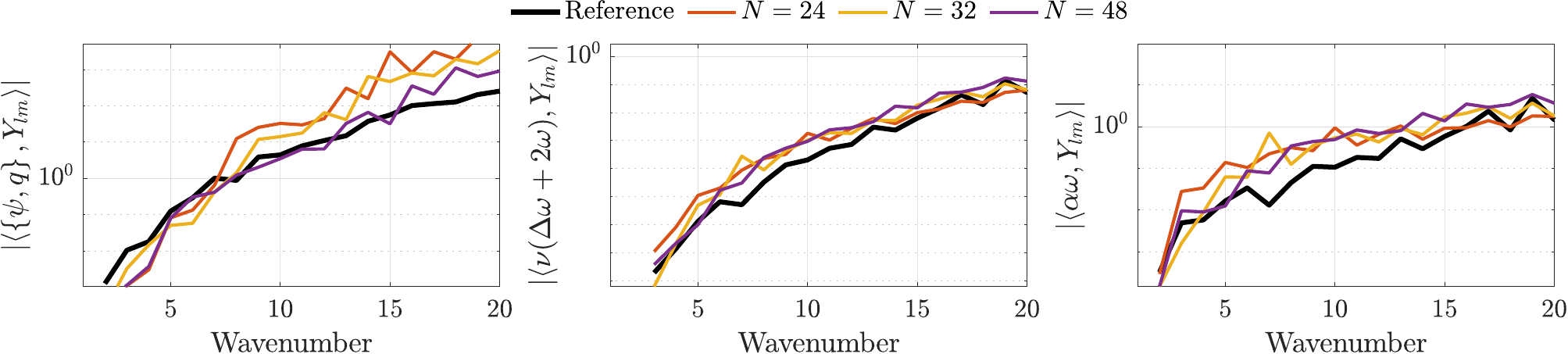}
    \caption{Instantaneous contributions per wavenumber for each of the terms in \eqref{eq:spectral_expansion} for the QGE with Lamb parameter $\gamma=10^4$. Shown are the contributions per time unit of the convective term (left), the diffusion terms (center), and the friction term (right). The quantities are measured 40 time units after initializing from the reference statistically stationary state.}
    \label{fig:lt_QG2_transfers}
\end{figure}

The results for the QGE are shown in Figures \ref{fig:lt_QG_vort_snaps} to \ref{fig:lt_QG2_transfers}. Comparing the instantaneous vorticity snapshots (Figures \ref{fig:lt_QG_vort_snaps} and \ref{fig:lt_QG2_vort_snaps}) shows that the zonal patterns in the vorticity are visible for the QGE with $\gamma=10^3$, but are difficult to identify when $\gamma=10^4$. Given the overall agreement in the energy spectra (Figures \ref{fig:lt_QG_vort_info} and \ref{fig:lt_QG2_vort_info}), the observed discrepancies are attributed to phase errors in the coarse model solutions. Similarly, the discrepancies between the reference solution and the model solutions, as far as diffusion and damping are concerned, are attributed to the qualitative differences in the instantaneous vorticity field. This is in contrast to the RNSE, where the latter two quantities were almost exactly reproduced in the coarse model simulations. Additionally, the zonal velocity is largely independent of the phases of the coefficients and is captured well. Extending the model construction and explicitly including the requirement to accurately predict the inter-scale energy transfer may further optimize the model parameters and reduce phase errors.

\subsection{Model predictions with partial mode corrections}
\label{subsec:partial_forcing}
We now turn our attention to the prediction quality when the closure model is applied to only a part of the modes. Reducing the number of forced modes reduces the number of model parameters and the required amount of data, which is beneficial when dealing with large data sets. Conversely, such reduced forcing also implies less control over nudging the flow toward its desired dynamics. A balance between these two requirements should be found. This is achieved by combining prior knowledge of the physical system with data-driven modeling and may be desired when aiming to reproduce a small number of flow statistics \cite{edeling2020reducing, bach2024filtering}. 

The ability of the closure model to reproduce select quantities with fewer forced modes is demonstrated in two numerical experiments. The results from Section \ref{subsec:known_ic} suggest that the most energetic modes of the solution are those approximately up to wavenumber $l=20$. Incidentally, the dominant zonal flow structures are represented by spherical harmonics with $m=0$. The first test performed here employs the model only for the most energetic modes, i.e., modes with $l=0, \ldots, 20$, $m=-l,\ldots, l$. In the second test, we exploit the known zonal structure by only applying the model to all modes $l=0,\ldots, N-1$ with $m=0$. Both experiments deal with the RNSE to clarify the presentation of the results. All other simulation parameters are the same as previously presented.

\begin{figure}
    \centering
    \includegraphics[width=\textwidth]{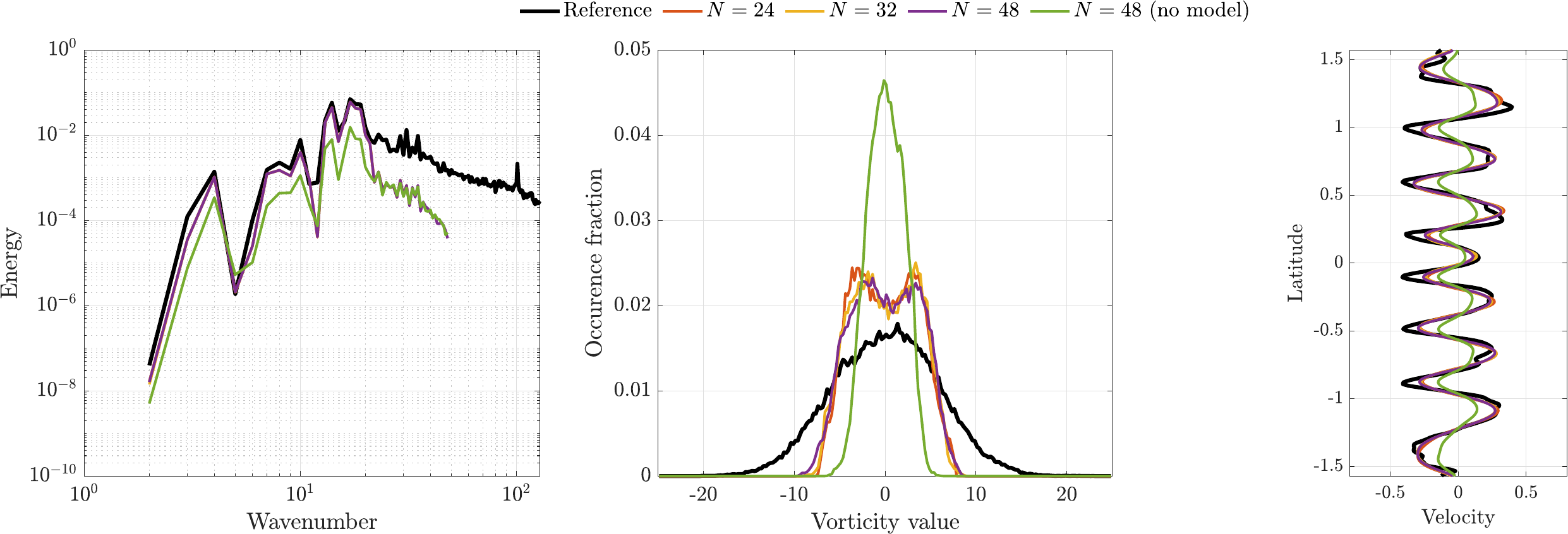}
    \caption{Instantaneous energy spectra (left), vorticity distribution (center), and zonal velocity (right), for the RNSE when only applying forcing to wavenumbers $l=0,\ldots, 20$. The vorticity fields are obtained 40 time units after initializing from the reference statistically stationary state.}
    \label{fig:lt_NS_partial}
    \centering
    \includegraphics[width=\textwidth]{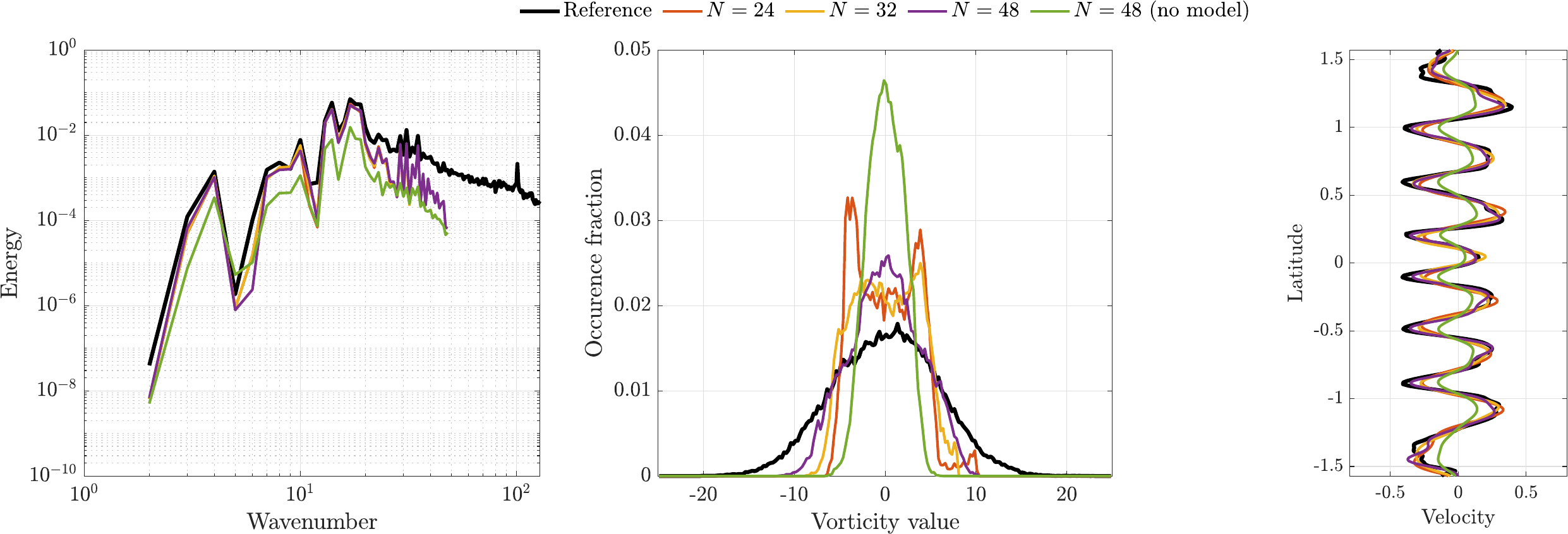}
    \caption{Instantaneous energy spectra (left), vorticity distribution (center), and zonal velocity (right), for the RNSE when only applying forcing to all resolvable zonal modes ($m=0$). The vorticity fields are obtained 40 time units after initializing from the reference statistically stationary state.}
    \label{fig:lt_NS_zonal}
\end{figure}

The results of the numerical tests are summarized in Figure \ref{fig:lt_NS_partial} and \ref{fig:lt_NS_zonal}, respectively. The results in Figure \ref{fig:lt_NS_partial} establish that the model reproduces the energy spectrum accurately up to the forced wavenumber. A qualitative deterioration of the vorticity distribution is observed compared to the results in Section \ref{subsec:known_ic}, visible as deviating shapes of the distributions. However, the zonal velocity is still maintained well despite the reduced forcing. Similar results are obtained for the second test, as depicted in Figure \ref{fig:lt_NS_zonal}. Forcing only the zonal modes approximates the reference spectrum reasonably well and maintains the zonal velocity accurately.

\newpage
\subsection{Model predictions from random initial conditions}
\label{subsec:random_ic}
We now assess the model performance after initializing the flow with a random vorticity field. In such cases, assimilating measured statistics in numerical simulations may accelerate convergence towards a statistically steady state \cite{bach2024filtering}. This can potentially be applied to reduce the spin-up time for numerical experiments of climate models \cite{bryan1984accelerating} and turbulence \cite{nelson2017reducing}. The purpose of the current tests is to assess the generalizability of the model for use on coarse grids. With random initial conditions, an accurate representation of the flow physics is not guaranteed when enforcing a select number of statistics  \cite{edeling2020reducing} and may help identify points of improvement of the model.

Initializing the flow from a random field outside the statistically steady state will yield coarse-resolution simulations that are unrecognizably different from the actual high-fidelity solution. Not resolving the external forcing leads to discrepancies even for the large-scale flow features, underlining the necessity of tailored explicit models to reproduce flow statistics. The initial conditions for each simulation are a random smooth vorticity field. Only the expansion coefficients of the zonal modes are real-valued, and we require that the initial signs of these coefficients in the coarse representation agree with the reference. The simulation parameters are otherwise as reported in Section \ref{subsec:governing_eqs}. The instantaneous snapshots of the vorticity are similar to those presented in Section \ref{subsec:known_ic} and are therefore omitted here. The results in this section are time averages, where the average is taken when the solutions have reached a statistically steady state. Each result is averaged over 200 consecutive flow snapshots separated by 0.1 time units.

\begin{figure}
    \centering
    \includegraphics[width=\textwidth]{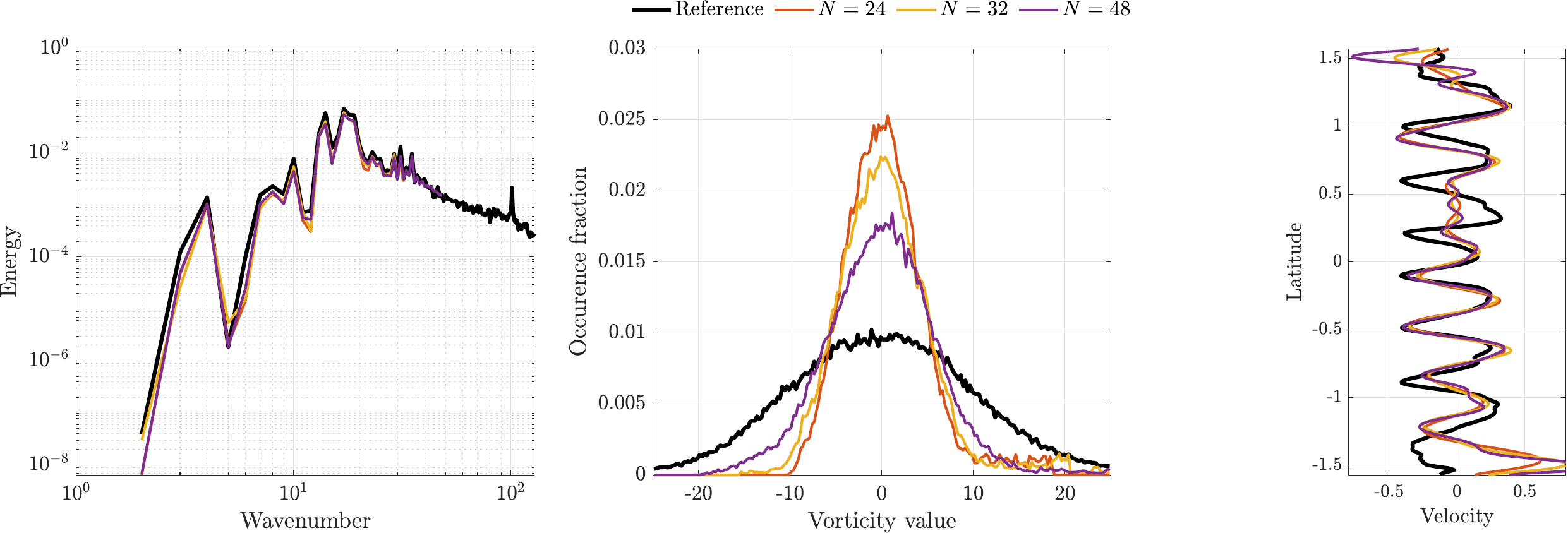} 
    \caption{Average energy spectra (left), vorticity distribution (middle), and zonal velocity (right) for the RNSE. The results are averaged over 200 snapshots.}
    \label{fig:NS_spectrum_zonalvel}
    \centering
    \includegraphics[width=\textwidth]{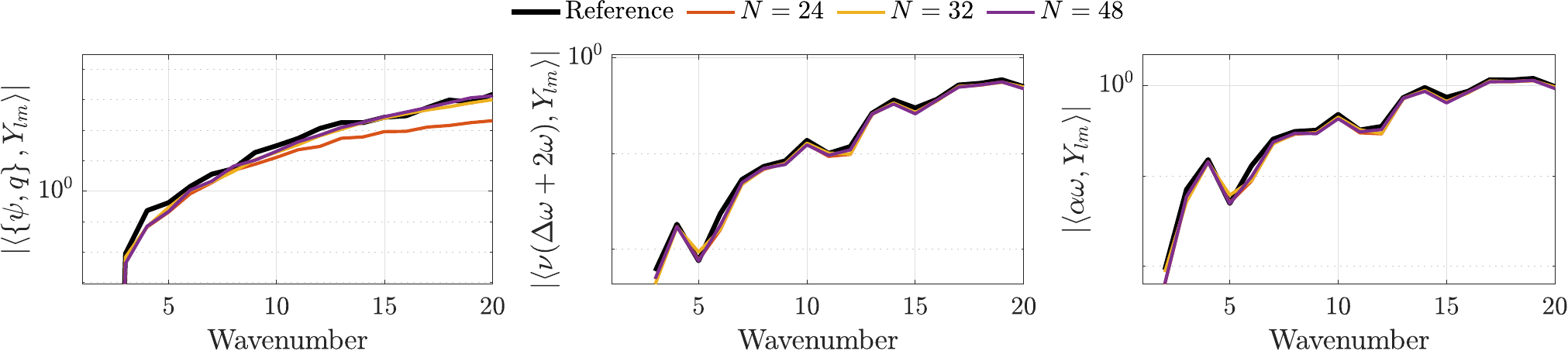}
    \caption{Average contributions per wavenumber for each of the terms in \eqref{eq:spectral_expansion} for the RNSE. Shown are the contributions per time unit of the convective term (left), the diffusion terms (center), and the friction term (right). The quantities are averaged over 200 snapshots.}
    \label{fig:NS_energy_transfer}
\end{figure}

The results for the RNSE are shown in Figures \ref{fig:NS_spectrum_zonalvel} and \ref{fig:NS_energy_transfer}. The energy spectra depicted in the left panel of Figure \ref{fig:NS_spectrum_zonalvel} indicate that the mean energy spectrum of the reference solution is accurately reproduced in all coarse numerical simulations, up to the smallest resolvable scales at the chosen resolution. The average zonal velocities are shown in the right panel of Figure \ref{fig:NS_spectrum_zonalvel} and illustrate that the reference velocity profiles can be accurately captured on all considered coarse grids.

Despite the qualitative agreement between the reference and coarse numerical solutions, deviations may be observed in the zonal velocity in Figure \ref{fig:NS_spectrum_zonalvel}. These deviations are localized near the poles and around $0.2-0.6$ in latitude. Comparison with the results in Section \ref{subsec:known_ic} suggests that this is caused by the random initial condition. In particular, we attribute this discrepancy to the expansion coefficients of the zonal modes having the wrong sign. The agreement of the coarse-grid energy spectra with the reference spectra only imposes a constraint on the magnitudes of the basis coefficients, but not on their phases. Despite including the correct coefficient signs in the initial condition, fully complying with the reference solution, the model does not explicitly enforce this correspondence during the subsequent numerical simulations. As a result, this appears to induce the observed discrepancies near the poles.

The average zonal velocity profiles exhibit a modest grid dependence. This illustrates that once the coarse grid forcing is obtained, the corresponding closure is quite independent of the adopted discretization method and resolution - this can be traced back to the adopted 3D-Var approach in which the $L_c$ and $L_r$ operators are assumed to be identity operators. Dependence on the adopted resolution is arguably a desirable feature of a closure model. Typically, in large-eddy simulation, the length-scale associated with the closure is taken as the mesh size~\cite{geurts2022direct}. This implies that increased spatial resolution diminishes the closure term \cite{geurts2002framework, geurts2002alpha}. Ultimately, adopting higher resolutions also decreases the effects of truncation and discretization errors, thereby decreasing the closure term and enabling approaching high-fidelity simulations as a consistency feature. Restoring some dependence on the discretization and resolution by extending the data assimilation algorithm may remove the grid-independence of the coarse-grid predictions. This is subject of ongoing research.

The average contribution of each of the terms in the dynamics is depicted in Fig. \ref{fig:NS_energy_transfer}. A comparison between the reference solution of the RNSE and the corresponding coarse-grid model solutions shows that the convective terms at $N=24$ deviate significantly from the reference. The magnitudes at lower wavenumbers ($l\leq 15$) are underestimated at resolutions $N=32$ and $N=48$. This indicates that more stringent constraints on the coefficient evolution may help improve the model.

\begin{figure}
    \centering
    \includegraphics[width=\textwidth]{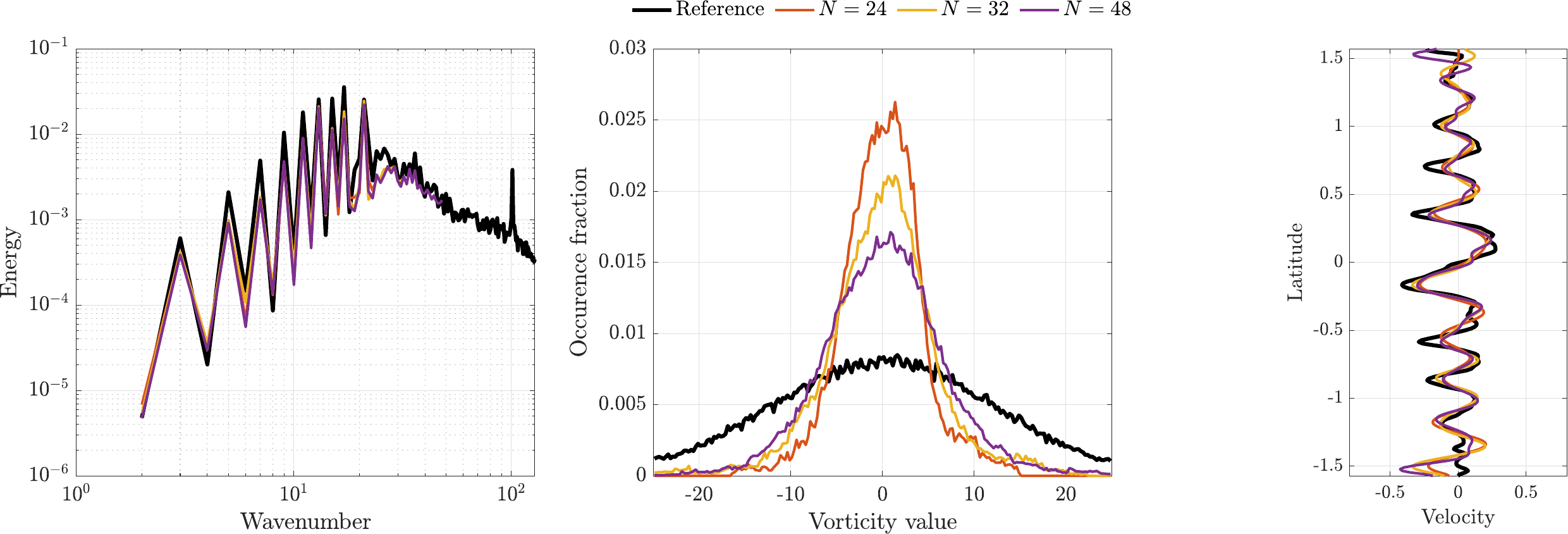} 
    \caption{Average energy spectra (left), vorticity distribution (middle), and zonal velocity (right) for the QGE with Lamb parameter $\gamma=10^3$. The results are averaged over 200 snapshots.}
    \label{fig:QG_spectrum_zonalvel}
    \centering
    \includegraphics[width=\textwidth]{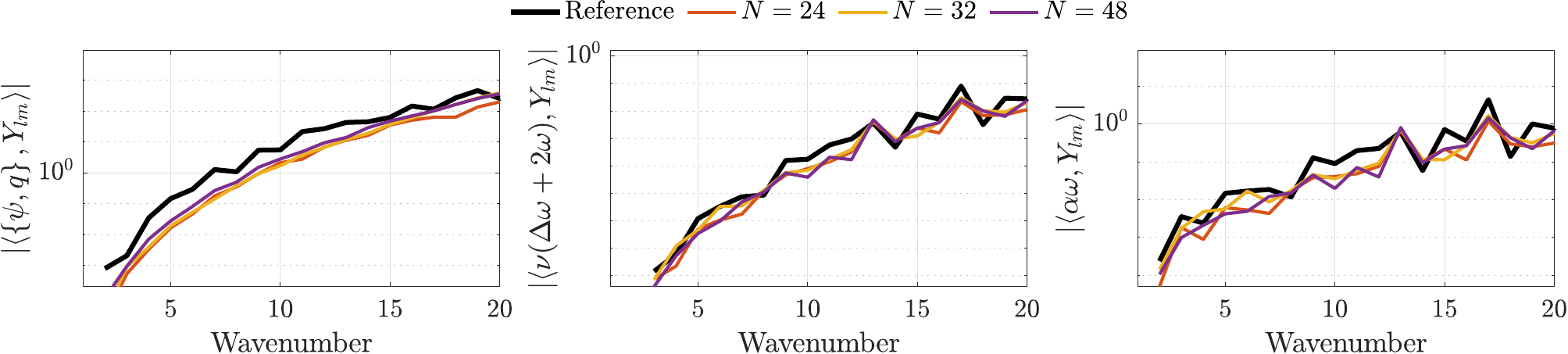}
    \caption{Average contributions per wavenumber for each of the terms in \eqref{eq:spectral_expansion} for the QGE with Lamb parameter $\gamma=10^3$. Shown are the contributions per time unit of the convective term (left), the diffusion terms (center), and the friction term (right). The quantities are averaged over 200 snapshots. }
    \label{fig:QG_energy_transfer}
\end{figure}

\begin{figure}
    \centering
    \includegraphics[width=\textwidth]{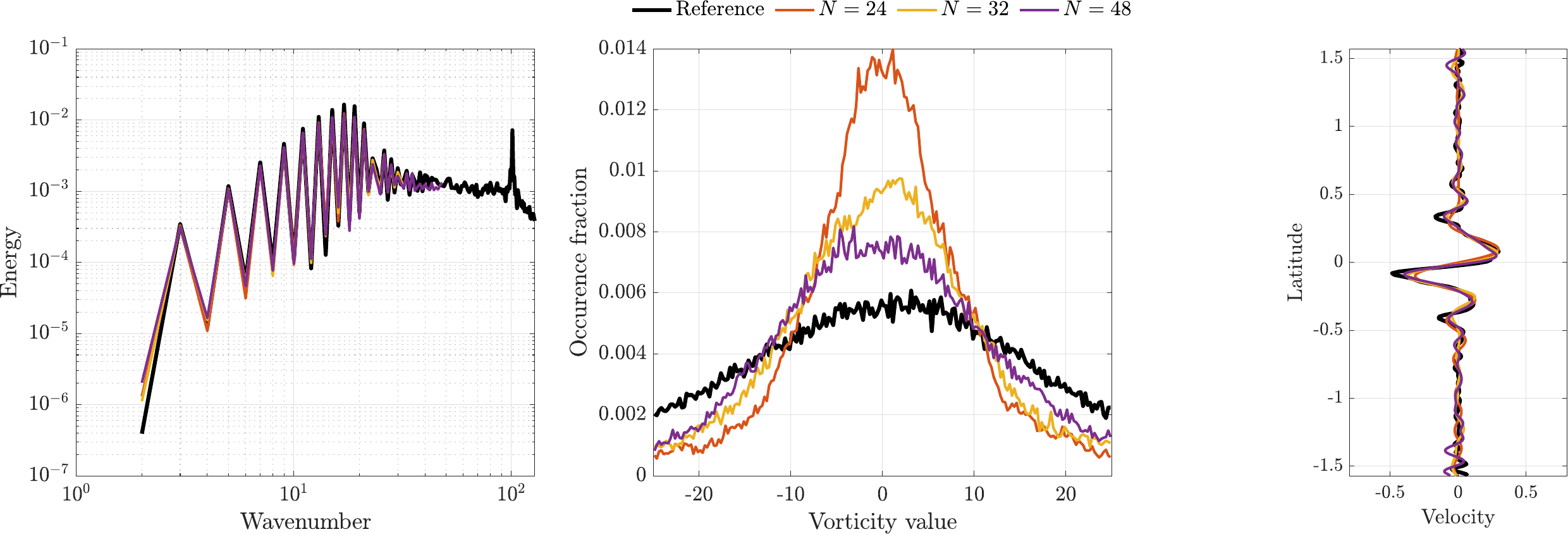}
    \caption{Average energy spectra (left), vorticity distribution (middle), and zonal velocity (right) for the QGE with Lamb parameter $\gamma=10^4$. The results are averaged over 200 snapshots.}
    \label{fig:QG2_spectrum_zonalvel}
    \includegraphics[width=\textwidth]{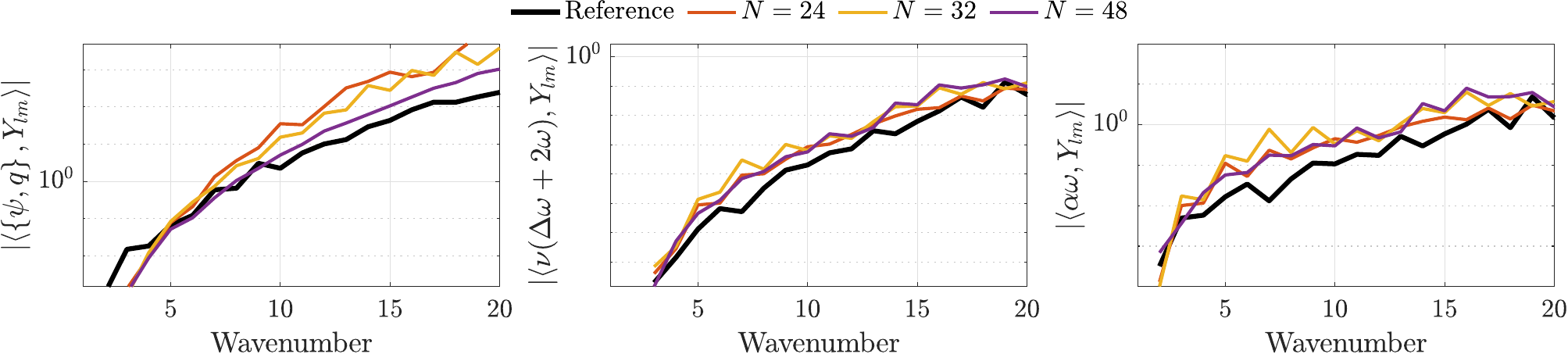}
    \caption{Average contributions per wavenumber for each of the terms in \eqref{eq:spectral_expansion} for the QGE with Lamb parameter $\gamma=10^4$. Shown are the contributions per time unit of the convective term (left), the diffusion terms (center), and the friction term (right). The quantities are averaged over 200 snapshots. }
    \label{fig:QG2_energy_transfer}
\end{figure}

The test cases for the QGE at $\gamma=10^3$ (Figures \ref{fig:QG_spectrum_zonalvel} and \ref{fig:QG_energy_transfer}) and $\gamma=10^4$ (Figures \ref{fig:QG2_spectrum_zonalvel} and \ref{fig:QG_energy_transfer}) display similar qualitative results as observed for the RNSE. The average energy spectra show good agreement up to the smallest resolvable scales, as designed, for both test cases at all adopted coarse resolutions. Good agreement is observed for the average zonal velocity, especially capturing the tapering profile in the latitudinal direction. Similar to the results of the RNSE, some deviations of the zonal velocity are observed near the poles which are again attributed to phase errors in the instantaneous vorticity fields. This reflects that a correct energy spectrum is a necessary but not sufficient modeling criterion. This suggests that imposing further model constraints might be desirable to improve predictions of these and higher-order moments, e.g., by employing statistical quantities such as the energy rate of change \cite{thuburn2014cascades, cifani2022sparse}. This is further highlighted in the comparison between the contributions of each term in the dynamics for the two QGE cases, as shown in Figures \ref{fig:QG_energy_transfer} and \ref{fig:QG2_energy_transfer}. The reference values of the magnitudes of the convective term are not followed as closely as observed for the RNSE. In particular, for $\gamma=10^3$ the magnitude of the convective terms is generally underestimated. For $\gamma=10^4$, these magnitudes are somewhat underestimated at the largest scales (up to wavenumbers 4) and overestimated at the smaller scales.

\begin{figure}
    
\end{figure}

\section{Conclusions}
\label{sec:conclusions}
In this work, a data-driven stochastic closure for turbulence modeling in large-eddy simulation (LES) was presented based on the 3D-Var data assimilation algorithm. The closure is motivated by the theoretical connection between so-called ideal LES and data assimilation. The added feedback forcing term is designed specifically to approximate the energy spectrum and is based on reference flow statistics obtained from offline high-resolution simulations. The corresponding closure model has few tunable parameters and the reduced computational costs enable fast computation of stochastic ensemble predictions for indefinite times. 

The proposed model was applied to three generic cases of geostrophic turbulence, described by the rotating Navier-Stokes equations and the quasi-geostrophic equations on the sphere. The closure was found to accurately recover the energy spectra on several coarse computational grids, establishing the desired spectrum-reconstructing property of the model. As a result, qualitative agreement was observed in the key flow statistics when applying the model after initializing the flow from known initial conditions. Initialization from random initial conditions yielded satisfactory results but indicated at the same time that more stringent constraints may be beneficial to obtain stand-alone models with a wider range of applicability.

Further model development based on data assimilation techniques is ongoing. Several extensions of the currently presented model can be investigated, including, e.g., explicitly coupling the Ornstein-Uhlenbeck processes using covariance estimates, employing Bayesian modeling to specify the forcing parameters, including additional statistical information such as inter-scale energy transfer in the nudging procedure, or explicitly taking into account discretization effects by employing an ensemble Kalman filtering approach.

\section*{Acknowledgments}
The authors would like to thank Darryl Holm, of the Department of Mathematics, Imperial College London, for his input in the context of the SPRESTO project, funded by the Dutch Science Foundation (NWO) in their TOP1 program. Computing budget was made available through the `Multiscale Modeling and Simulation' project, supported by NWO, the National Science Foundation in the Netherlands. Simulations were executed on Snellius, the national supercomputing facility at SurfSara.

%\bibliographystyle{plain}
%\bibliography{biblio}% Produces the bibliography via BibTeX.
\printbibliography

\end{document}